\documentclass{iau379}

\usepackage{natbib}
\usepackage{graphicx}
\usepackage{amsmath}
\usepackage{multirow}
\usepackage{journals}

\begin{document}

\lefttitle{Chilingarian et al.}
\righttitle{Dwarf early-type galaxies in the Coma cluster}

\jnlPage{1}{7}
\jnlDoiYr{2023}
\doival{10.1017/xxxxx}

\aopheadtitle{Proceedings of IAU Symposium 379}
\editors{P. Bonifacio,  M.-R. Cioni, F. Hammer, M. Pawlowski, and S. Taibi, eds.}

\title{Dwarf early-type galaxies in the Coma cluster: internal dynamics, stellar populations}

\author{Igor V. Chilingarian$^{1,2}$, Kirill A. Grishin$^{3,2}$, Anton V. Afanasiev$^{4,2}$,
Anton Mironov$^{5}$, Daniel Fabricant$^{1}$, Sean Moran$^{1}$, Nelson Caldwell$^{1}$, Ivan Yu. Katkov$^{6,2}$, Irina Ershova$^{5}$}

\affiliation{$^1$Center for Astrophysics -- Harvard and Smithsonian, 60 Garden St. Cambridge, MA, 02138 USA\\
$^2$Sternberg Astronomical Institute, Moscow State University, 13 Universitetsky prospect, Moscow, Russia\\
$^3$Universit\'e Paris Cit\'e, CNRS, Astroparticule et Cosmologie, F-75013 Paris, France\\
$^4$LESIA, Observatoire de Paris, 5 place Jules Janssen, 92195, Meudon, France\\
$^5$Faculty of Space Research, Moscow State University, 1 Leninskie Gory, bld.~52, Moscow, Russia\\
$^6$New York University Abu Dhabi, Saadiyat Island, PO Box 129188, Abu Dhabi, UAE}

\begin{abstract}
We present preliminary results from our spectroscopic survey of low-luminosity early-type galaxies in the Coma cluster conducted with the Binospec spectrograph at the 6.5~m MMT. From spatially-resolved profiles of internal kinematics and stellar population properties complemented with high-resolution images, we placed several low-luminosity dEs on the fundamental plane in the low-luminosity extension of the available literature data. We also discovered unusual kpc-sized kinematically-decoupled cores in several dwarf galaxies, which had been probably formed before these galaxies entered the cluster.
\end{abstract}

\maketitle

\section{Introduction}

Dwarf elliptical (dE) and lenticular (dS0) galaxies are low-stellar-mass ($M_*\lesssim 5\cdot10^9 M_{\odot}$) quiescent stellar systems commonly found in galaxy clusters and massive groups \citep{SB84}. The formation and evolution of dE/dS0s are thought to be influenced by a combination of internal processes, such as stellar feedback \citep{1986ApJ...303...39D} and environmental effects, ram pressure stripping of gas by the hot intracluster medium \citep{1972ApJ...176....1G} and tidal interactions with other cluster members as well as the cluster potential \citep{1996Natur.379..613M}. Dwarf early-type galaxies are not as simple as they look. Faint low-contrast spiral arms and bars were found in several `luminous' dE/dS0s in the Virgo and Abell~496 clusters \citep{2000A&A...358..845J,2008A&A...486...85C}. The diversity of rotational support in dE/dS0s \citep{2002AJ....124.3073G}, residual central star formation \citep{2006AJ....132.2432L}, and young metal-rich embedded substructures  and kinematically decoupled cores \citep{CPSA07,2007AstL...33..292C,2009MNRAS.394.1229C} strengthened their evolutionary connection to more massive `normal' galaxies \citep{2008ApJ...674..742B}.
However, which evolutionary mechanism dominates the dE/dS0 formation is still not understood. In 2017--2022 we carried out a spectroscopic campaign to characterise dE/dS0 populations in three nearby clusters ($0.023<z<0.045$) and assess their evolutionary paths by obtaining spatially resolved internal kinematics and stellar population characteristics of several hundred dE/dS0 galaxies.

\section{Observations, Data Reduction and Analysis}
We observed galaxies in the three low-redshift clusters, Coma$=$Abell~1656 ($z=0.023$), Abell~2147 ($z=0.035$), and Abell~168 ($z=0.045$) using the high-throughput multi-object optical spectrograph Binospec \citep{2019PASP..131g5004F} operated at the f/5 focus of the 6.5~m MMT at Mt.~Hopkins, Arizona. The field-of-view of 16'$\times$15' at this redshift range allows us to observe up-to 100 galaxies in each multi-slit mask at the spatial resolution of 0.4--0.7~kpc (corresponding to the typical 0.8'' seeing). Our sample includes $>200$ dEs/dS0s in the Coma cluster (8 masks), 30 in Abell~2147 (1 mask), and 50 in Abell~168 (2 masks). Using tilted slits, we observed a subsample of $\sim100$ Coma and Abell~168 dEs/dS0s along their major and minor axes. We integrated for 2--4~h on each mask and reached a surface brightness $\mu_B \sim 25.5$~mag~arcsec$^{-2}$ at S/N=3~pix$^{-1}$ in the wavelength range $3700<\lambda<5300$~\AA. The 1000~gpm grating provides a spectral resolution $R=4800$. This setup is well suited for internal kinematic studies and stellar population analysis of faint absorption-line spectra \citep{2020PASP..132f4503C,2023MNRAS.520.6312A}. In addition to the spectroscopic data, we used high-resolution Hubble Space Telescope images.

We reduced spectroscopic data using the Binospec data reduction pipeline \citep{2019PASP..131g5005K} optimized for low-surface brightness data. We used the {\sc NBursts} full spectum fitting code \citep{CPSK07} with $R=10000$ simple stellar population models computed with the {\sc pegase.hr} code \citep{LeBorgne+04} and a dedicated grid of synthetic spectra, which model a process of a ram-pressure-induced starburst \citep{2019arXiv190913460G} for a subsample of objects with signs of recent quenching. We analyzed images using {\sc galfit} \citep{2010AJ....139.2097P} by using up-to three light profile components and computed an equivalent half-light radius from several components. %Then we ran Jeans axisymmetric modelling which included a spherical dark matter halo in addition to the stellar component \citep{2023MNRAS.520.6312A}.

\section{Preliminary Results}
\subsection{Coma dEs in $\kappa$-space}
\begin{figure}
    \centering
    \includegraphics[width=0.115\hsize]{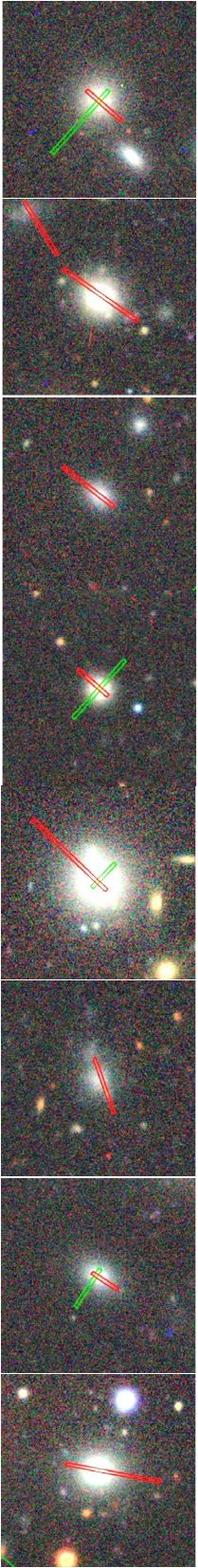} % 0.092\hsize
    \includegraphics[width=0.875\hsize]{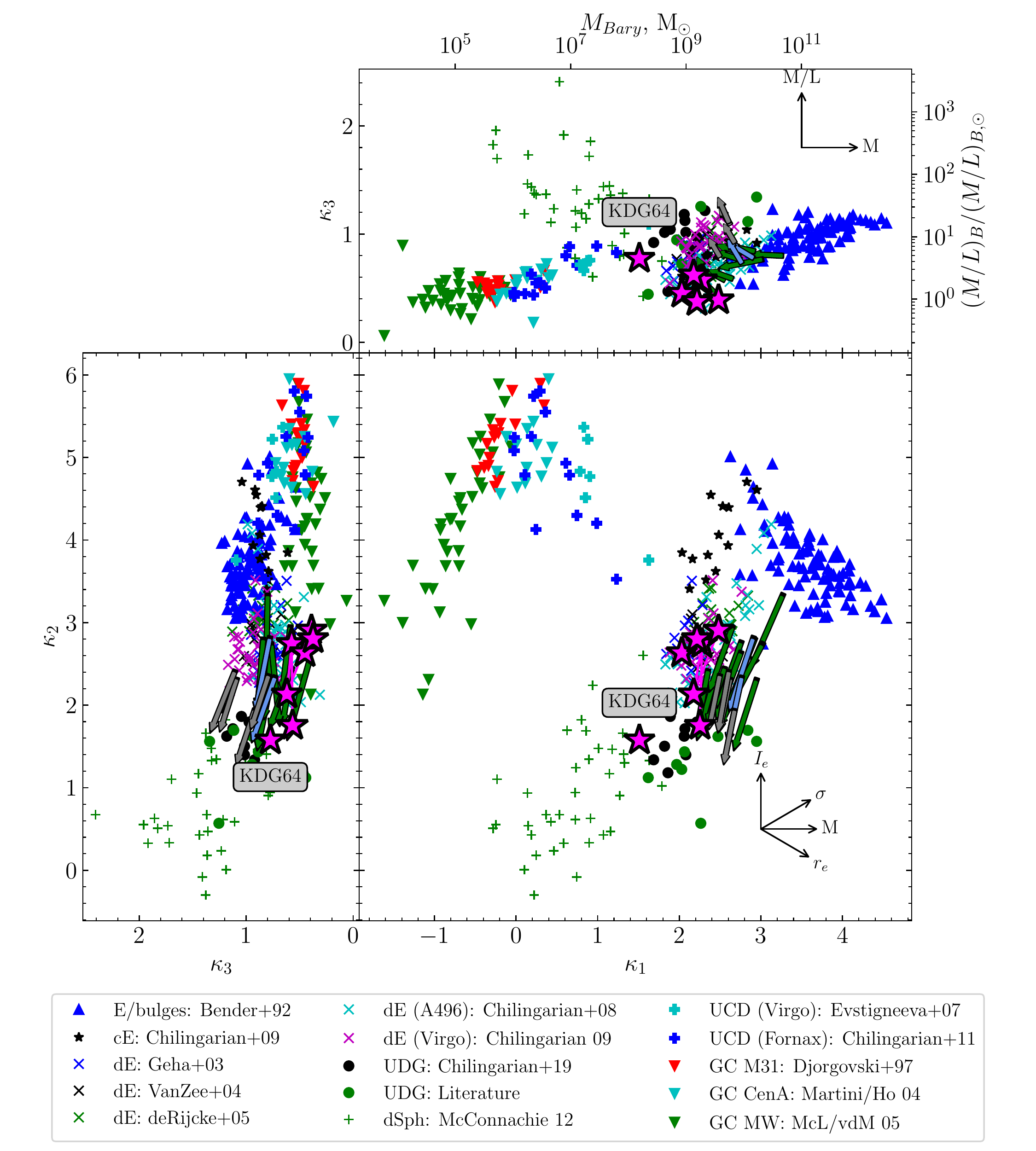} % 0.700\hsize
    \caption{\textbf{Left:} examples of Coma cluster dEs with positions of Binospec slits overplotted. Four galaxies were observed along two axes (green and red slits). \textbf{Right:} $\kappa$-space FP projection, where the results of our analysis for Coma cluster dEs are shown by purple stars. KDG~64, a large dwarf spheroidal or a small UDG in the M~81 group \citep{2023MNRAS.520.6312A} is marked: its position is clearly distinct from the remaining dEs. The arrow represent predicted passive evolution of diffuse post-starburst galaxies.}
    \label{fig:fp}
\end{figure}
A good proxy for a dynamical mass of a virialized stellar system is its position in the Fundamental Plane \citep[FP,][]{1987ApJ...313...59D}, which requires the knowledge of $R_e$, $\langle \mu_e \rangle$ and $\sigma_*$. Here we use a so-called $\kappa$-space FP projection \citep{1992ApJ...399..462B}, where one of the axes, $\kappa_3$, is related to the dynamical mass-to-light ratio and another one, $\kappa_1$ can be expressed as a logarithm of the stellar mass. In Fig.~\ref{fig:fp} we present a literature compilation of structural and dynamical measurements for dwarf and giant early-type galaxies, ultradiffuse galaxies (UDGs), bulges, and compact stellar systems. \nocite{2005A&A...438..491D} \nocite{1992ApJ...399..462B} \nocite{2004AJ....128.2797V} \nocite{2003AJ....126.1794G} \nocite{Chilingarian+08} \nocite{2005ApJS..161..304M} \nocite{1997ApJ...474L..19D} \nocite{2004ApJ...610..233M} \nocite{2012AJ....144....4M} \nocite{Chilingarian+09} \nocite{Chilingarian+19} \nocite{2007AJ....133.1722E} \nocite{Chilingarian+11} \nocite{2009MNRAS.394.1229C} \nocite{2021NatAs...5.1308G} New measurements from Binospec spectra and HST images are shown as purple stars. The purple stars fall on the extension of the dE locus towards lower stellar masses and lower velocity dispersions. Interestingly, these objects seem to have lower dynamical mass-to-light ratios than Virgo cluster dEs of the same stellar mass. All fall below the $M/L$ rise in the dwarf spheroidal regime, where KDG~64, one of the most massive dSphs in the Local Volume, resides \citep{2023MNRAS.520.6312A}. Interestingly, UDGs have higher $M/L$ ratios for the same stellar mass.

\subsection{Kinematically decoupled cores}
\begin{figure}
    \centering
    \includegraphics[width=\hsize]{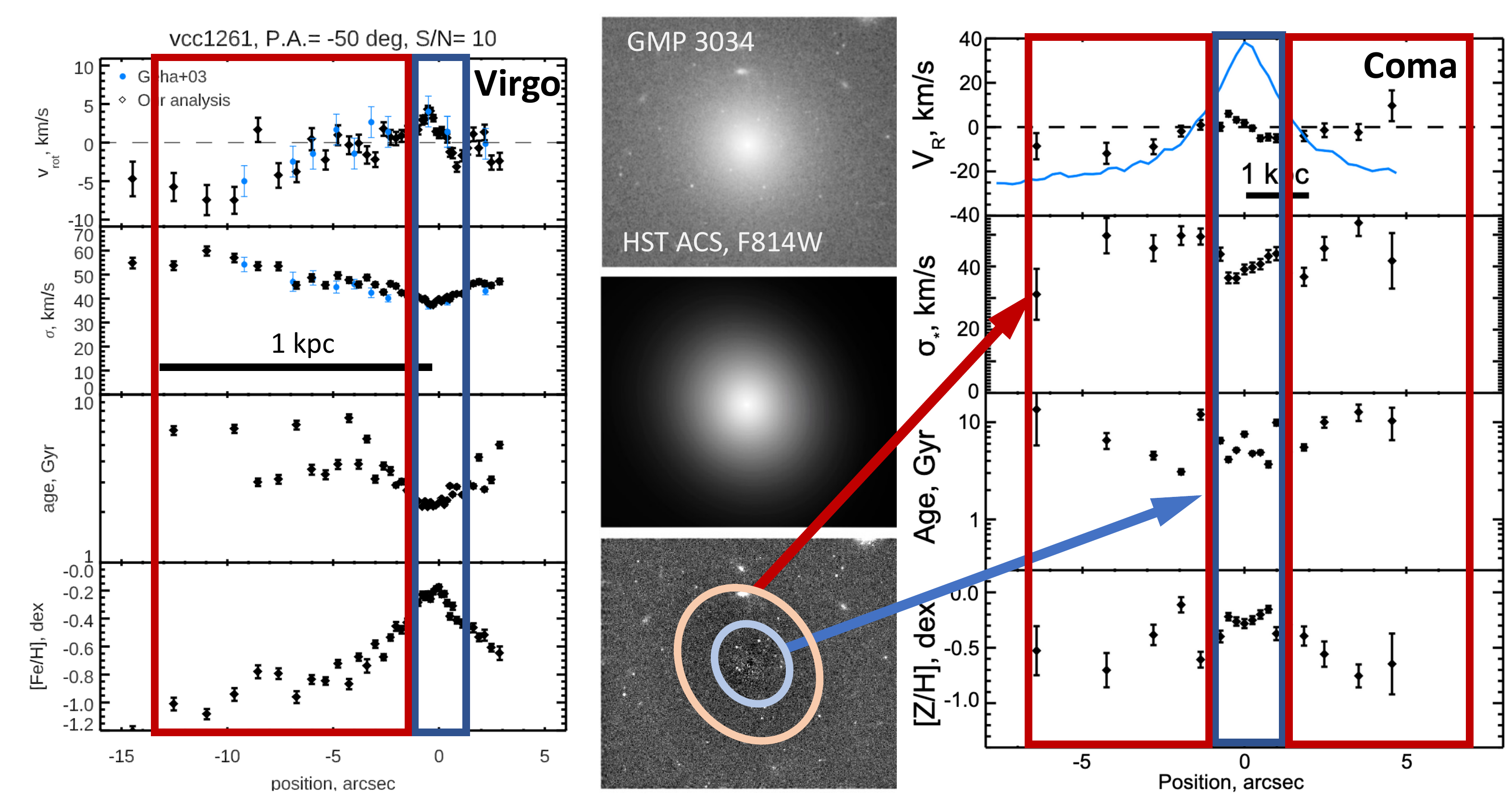}
    \caption{Examples of kinematically decoupled cores found in VCC~1261 in the Virgo cluster (re-analysis of archival observations from Keck ESI presented in \citealp{2003AJ....126.1794G}) and GMP~3034 (Binospec data). Note the difference of spatial scales (0.08 and 0.5~kpc~arcsec$^{-2}$ for Virgo and Coma respectively).}
    \label{fig:kdc}
\end{figure}

Another intriguing result of our project is a discovery of several kinematically decoupled cores (KDCs) in brighter ($M_V<-17$~mag) Coma cluster dEs. In the Coma cluster, to our surprise, we found KDCs which are much more extended ($>1$~kpc) than those in the Virgo cluster (see Fig.~\ref{fig:kdc}). Because the distance is 6 times greater, we cannot detect KDCs as compact as those in the Virgo cluster ($200\dots300$~pc) in ground-based seeing-limited observations, however, no `large' KDCs have been found in the Virgo dEs to date \citep{2014ApJ...783..120T}. Both `small' KDCs in Virgo dEs and `large' KDCs in Coma dEs are usually distinct in stellar age and metallicity; they are younger and more metal-rich. KDCs are thought to form via minor mergers, cold gas accretion, or galaxy flybys at a low relative velocity. While mergers can happen in clusters, despite a low probability of dwarf-dwarf interactions, the two latter channels are practically unfeasible due to the dynamically hot environment. However, a pre-existing KDC in a disk galaxy can survive its morphological transformation into a dE during infall. The environmental phenomena in the massive Coma cluster are expected to be more violent than in Virgo, therefore more massive galaxies with larger pre-existing KDCs can be efficiently quenched and transformed into `oversized' dEs with large KDCs.

\bibliographystyle{aasjournal_nodoi}
\bibliography{dE2023.bib}

%\begin{discussion}
%\discuss{Fourth}{Why do you believe proceedings are useful ?}
%\discuss{Someone}{Because they allow to capture the Symposium and provide an updated status of the research field.}
%\end{discussion}

\end{document}